\documentclass[sigconf]{acmart}
\acmConference[MSR 2026]{MSR '26: Proceedings of the 23rd International Conference on Mining Software Repositories}{April 2026}{Rio de Janeiro, Brazil}

\AtBeginDocument{%
  }

\copyrightyear{2026}
\acmYear{2026}
\setcopyright{cc}
\setcctype{by}
\acmConference[MSR '26]{23rd International Conference on Mining Software Repositories}{April 13--14, 2026}{Rio de Janeiro, Brazil}
\acmBooktitle{23rd International Conference on Mining Software Repositories (MSR '26), April 13--14, 2026, Rio de Janeiro, Brazil}
\acmPrice{}
\acmDOI{10.1145/3793302.3793615}
\acmISBN{979-8-4007-2474-9/2026/04}

\usepackage{tabularx}
\usepackage{subcaption}
\usepackage{tabularx}
\usepackage[dvipsnames]{xcolor}
\usepackage[normalem]{ulem}
\usepackage{enumitem}
\usepackage{pifont}
\usepackage{pgfplots}
\usepackage{array}
\usepackage{lipsum}
\usepackage[utf8]{inputenc}
\usepackage{hyperref}
\usepackage{tabularray} 
\usepackage{graphicx}
\usepackage{caption}
\usepackage{tikz}
\usetikzlibrary{pgfplots.statistics, pgfplots.colorbrewer} 
\usepackage{ragged2e}
\usepackage{makecell}

\usepackage[most]{tcolorbox}
\usetikzlibrary{patterns}

\UseTblrLibrary{booktabs} 
\usetikzlibrary{calc}

\hypersetup{
    colorlinks=true,
    linkcolor=blue,
    filecolor=magenta,      
    urlcolor=cyan,
    pdftitle={Overleaf Example},
    pdfpagemode=FullScreen,
    }
\usepackage{amsmath}
\usepackage{multicol}
\usepackage{multirow}
\definecolor{Mycolor}{HTML}{166666}
\definecolor{mygrayzero}{gray}{0.9}
\definecolor{mygrayone}{gray}{0.8}
\definecolor{mygraytwo}{gray}{0.7}
\definecolor{mygraythree}{gray}{0.6}

\definecolor{Mycolor}{HTML}{166666}
\newboolean{showcomments}
\setboolean{showcomments}{true}
\ifthenelse{\boolean{showcomments}}
{
  \newcommand{\nbc}[3]{
    \colorbox{#3}{\bfseries\sffamily\scriptsize\textcolor{white}{#1}}
    {\textcolor{#3}{\sf\small$\blacktriangleright$\textit{#2}$\blacktriangleleft$}}
  }
}
{
  \newcommand{\nbc}[3]{}
}

\begin{document}

\title{Beyond Bug Fixes: An Empirical Investigation of Post-Merge Code Quality Issues in Agent-Generated Pull Requests}

\author{Shamse Tasnim Cynthia} 
\affiliation{
    \institution{University of Saskatchewan} 
    \city{Saskatoon}
    \country{Canada}}
\email{shamse.cynthia@usask.ca}
\author{Al Muttakin}
\affiliation{
    \institution{University of Saskatchewan} 
    \city{Saskatoon}
    \country{Canada}}
\email{al.muttakin@usask.ca}
\author{Banani Roy}
\affiliation{
    \institution{University of Saskatchewan} 
    \city{Saskatoon}
    \country{Canada}}
\email{banani.roy@usask.ca}

\renewcommand{\shortauthors}{Trovato et al.}

\begin{abstract}
The increasing adoption of AI coding agents has increased the number of agent-generated pull requests (PRs) merged with little or no human intervention. Although such PRs promise productivity gains, their post-merge code quality remains underexplored, as prior work has largely relied on benchmarks and controlled tasks rather than large-scale post-merge analyses. To address this gap, we analyze 1,210 merged agent-generated bug-fix PRs from Python repositories in the AIDev dataset. Using SonarQube, we perform a differential analysis between base and merged commits to identify code quality issues newly introduced by PR changes. We examine issue frequency, density, severity, and rule-level prevalence across five agents. Our results show that apparent differences in raw issue counts across agents largely disappear after normalizing by code churn, indicating that higher issue counts are primarily driven by larger PRs. Across all agents, code smells dominate, particularly at critical and major severities, while bugs are less frequent but often severe. 
Overall, our findings show that merge success does not reliably reflect post-merge code quality, highlighting the need for systematic quality checks for agent-generated bug-fix PRs.

\end{abstract}

\ccsdesc[500]{Software and its engineering~Maintaining software}

\keywords{Coding Agents, Code Quality, Static Analysis, Bug-fix, SonarQube}

\maketitle

\section{Introduction} \label{Sec:Introduction}
Software Development Agents (e.g., Codex, Copilot) represent the next step in AI-powered development by accelerating the development lifecycle \cite{hong2024metagpt, qian-etal-2024-chatdev}.
Through reasoning, planning, and environmental interaction, these agents perform complex coding tasks and contribute autonomously to software repositories \cite{DBLP:journals/tosem/HouZLYWLLLGW24, horikawa2025agentic}. However, recent evidence shows that agentic pull requests (PRs) are less likely to be merged, particularly those involving bug fixes or feature development \cite{li2025aidev}. This raises concerns about the quality risks that may accompany the integration of agentic changes. These concerns are particularly important for bug-fix PRs, which are expected to improve the correctness and reliability of the codebase. Thus, a systematic investigation of agentic bug-fix PRs is warranted to assess whether they introduce post-merge code quality issues that could affect maintainability, reliability, or security of the software system.


Prior studies have evaluated AI coding assistants using controlled settings and benchmark-based tasks. Nikolaidis et al.~\cite{nikolaidis2024chatgptcopilot} compared ChatGPT and GitHub Copilot on Python tasks and found that, while solutions are often runnable, they frequently contain logical errors and require refinement. However, such problem-solving benchmarks do not capture PR-based integration, where changes interact with evolving codebases and are merged into production. Chen et al.~\cite{chen2025evaluating} evaluated ten coding agents on real-world GitHub issues from SWE-bench, but these benchmarks still fail to reveal large-scale post-merge quality trends across agents and repositories. As a result, we still lack a large-scale characterization of the quality issues introduced by merged agentic bug-fix PRs.

In this study, we analyzed 1,210 merged agentic bug-fix PRs from 206 Python repositories in the AIDev dataset \cite{li2025aidev}. First, we assess the post-merge code quality using SonarQube \cite{sonarqube_product_2025} through a differential analysis that compares each PR’s base (parent) commit and merged commit to identify issues newly introduced by the PR changes. Second, we characterize introduced issue types(e.g., bugs, code smells, vulnerabilities, security hotspots, duplication indicators, and added technical debt) and analyze their distribution across agents. Finally, we examine the severity profile of the introduced issues and identify the most frequently occurring violated rules across different agent types, highlighting recurring patterns that may represent systematic risks in agentic bug-fix contributions. Throughout, we treat SonarQube outputs as indicators of quality risk and report descriptive post-merge evidence.
by answering the following two research questions.
\begin{itemize}[leftmargin=*] 
 \renewcommand\labelitemi{\ding{43}}
    \item \textbf{RQ\textsubscript{1}.}~\textbf{To what extent are code quality issues present in merged Fix-PRs and how do observed patterns vary across agents?}
    \item \textbf{RQ\textsubscript{2}.}~\textbf{What is the severity profile of code quality issues introduced by merged PRs, and which issues occur most frequently across different agent types?} 
\end{itemize}

\textbf{Replication Package }can be found in our online appendix \cite{replication_package}.

\section{Methodology} \label{Sec:Methodology}
The subsequent sections provide the details of our methodology.
\textbf{Dataset Construction:}
We use the AIDev \cite{li2025aidev}, a large-scale collection of agent-authored PRs mined from GitHub repositories. We start from a curated subset of AIDev that contains $33{,}596$ agentic PRs from $2{,}807$ repositories with more than 100 stars, categorized by task type and authored by five agents (OpenAI Codex, Copilot, Devin, Cursor, and Claude Code). From this subset, we first select the $8{,}106$ \textit{fix}-type (bug remediation) PRs. Next, we restrict the scope to Python repositories and extracted $1{,}802$ PRs from $206$ Python projects. Finally, from these PRs, we filtered out $1{,}210$ that were merged and use them for the subsequent analysis. We intentionally focus on bug-fix PRs because their primary intent is to remediate known defects; therefore, it is especially important to examine whether such changes introduce additional quality issues after integration \cite{Hanna2023HotPHA}. Second, we limit the analysis to Python repositories to ensure methodological consistency and keep the study tractable while preserving internal validity. Python is also widely used in modern software development and strongly represented in AI-assisted coding workflows, making it a practical and relevant scope for studying agentic PRs. Table~\ref{tab:dataset-summary} summarizes the dataset used in this study. As shown in the Table, the distribution of agent-generated PRs in our dataset is imbalanced, with OpenAI Codex accounting for the majority of merged PRs.
\begin{table}[H]
  \centering
  \vspace{-0.3cm}
  \caption{An overview of the subject dataset}
  \vspace{-0.3cm}
  \label{tab:dataset-summary}
  \resizebox{0.6\linewidth}{!}{
      \begin{tabular}{lr}
        \toprule
        \textbf{Item} & \textbf{Value}\\
        \midrule
        \# of total \textit{fix}-type agentic PRs            & 8,106 \\
        \# of merged PRs (Python repositories)               & 1,210 \\
        \midrule
        \# of OpenAI Codex PRs                               & 949 \\
        \# of Copilot PRs                                    & 106 \\
        \# of Devin PRs                                      & 100 \\
        \# of Cursor PRs                                     & 40 \\
        \# of Claude Code PRs                                & 15 \\
        \bottomrule
      \end{tabular}
    }
\end{table}
\vspace{-0.3cm}

\textbf{SonarQube analysis for Agentic PRs}
To measure the code quality issues introduced by each PR, we use SonarQube \cite{sonarqube_product_2025} on the $1{,}210$ PRs from our dataset and perform a differential analysis that contrasts the repository state before and after the PR is merged. SonarQube is widely adopted in both industry and research, and has been used in numerous prior studies \cite{dantas2023developers, chen2025evaluating}. It reports a broad range of code quality concerns—including bugs, code smells, security hotspots, and violations of predefined coding rules—across multiple programming languages. For each PR in our dataset, we check out the \textit{base (parent)} commit representing the code state before the merge and run SonarQube analysis. We then check out the \textit{merged} commit and rerun SonarQube under the same configuration to identify newly introduced issues merged PR.

\textbf{Frequency \& Distribution (RQ\textsubscript{1}):}
After running the SonarQube analysis, we measure post-merge quality issues using SonarQube’s issue types and related quality dimensions. Specifically, we extract \textit{Bugs}, \textit{Code Smells}, \textit{Vulnerabilities}, and \textit{Security Hotspots} along with duplicate code blocks and the added technical debt (remediation effort) reported by SonarQube. We quantify issue presence and frequency at the PR level, and then aggregate these outcomes to characterize how issue types are distributed across different agents and projects. All SonarQube outputs are stored in a structured format to enable consistent aggregation and subsequent statistical analysis.
Next, to account for differences in PR size, we compute \textit{issue density} as the number of SonarQube issues normalized by the lines of code modified. For each PR, we extract the total lines added and deleted in the merged commit \cite{nagappan2005use, mcintosh2014impact} and compute issue density as issues per thousand lines of code (KLOC). This normalization supports fair comparisons across agents by controlling for variation in PR size.

\textbf{Severity \& Prevalence (RQ\textsubscript{2}):}
We further extract issue-level attributes to assess practical risk and recurring patterns. For each newly introduced issue, we record its SonarQube severity (INFO, MINOR, MAJOR, CRITICAL, BLOCKER) and the violated rule identifier. This enables comparisons of severity profiles across agent types and helps identify the most frequently violated rules overall for each agent. For Security Hotspots, we additionally record the hotspot risk/probability level (e.g., LOW, MEDIUM, HIGH), when available. Together, these attributes support both distributional comparisons and frequency-based ranking of recurring issue patterns.

\begin{table}[t]
\centering
\caption{Statistical Analysis of Issues by AI Coding Agent}
\label{tab:agent_statistics}
\resizebox{\linewidth}{!}{%
    \begin{tabular}{lrrrrrrrrrrr}
    \hline
    \textbf{Agent} & \textbf{\% w/ Issues} & \makecell{\textbf{Total}\\\textbf{Issues}} & \textbf{Mean} & \textbf{Median} & \makecell{\textbf{Std.}\\\textbf{Dev}} & \textbf{Max} & \makecell{\textbf{Total}\\\textbf{Debt (Hr)}}\\
    \hline
    Claude          & 53.3(8/15)       & 69    & 4.6 & 1.0 & 9.6 & 35 & 8.9\\
    Copilot         & 20.8(22/106)     & 235   & 2.2 & 0.0 & 10.1 & 94& 45.8\\
    Cursor          & 22.5(9/40)       & 331   & 8.3 & 0.0 & 38.2 & 233& 39.5\\
    Devin           & 26.0(26/100)     & 99    & 1.0 & 0.0 & 2.8 & 22 & 9.9\\
    Codex           & 13.2(125/949)    & 456   & 0.5 & 0.00 & 2.6 & 30& 41.2\\
    \hline
    \end{tabular}%
}
\vspace{-0.3cm}
\end{table}
\begin{figure}[t]
    \centering
    \resizebox{0.8\linewidth}{!}{
    \includegraphics[width=0.7\linewidth]{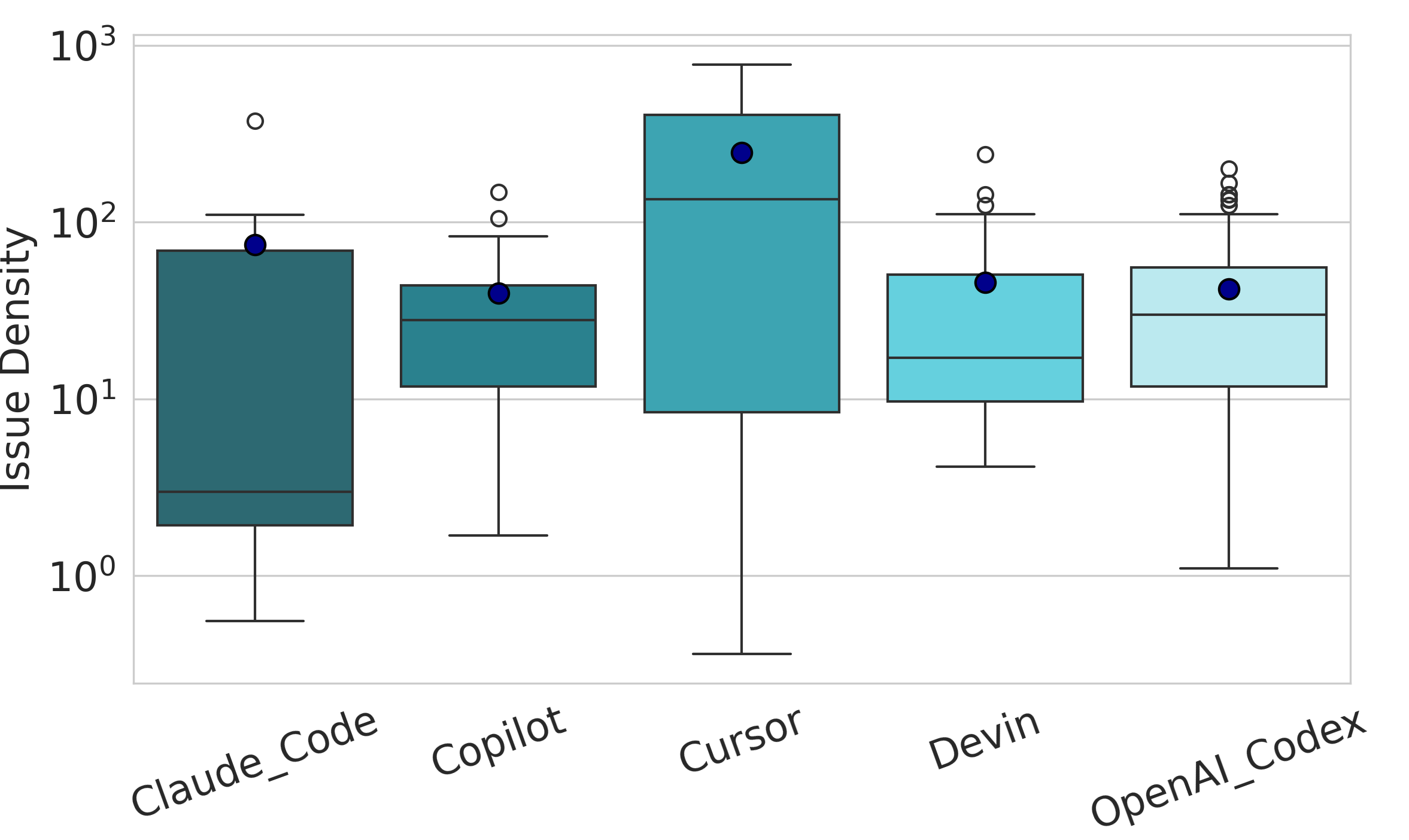}
    }
    \vspace{-0.4cm}
    \caption{Issue density across agents}
    \label{fig:issue_density}
    \vspace{-0.5cm}
\end{figure}
%
\section{Study Findings} \label{Sec:Study-Findings}

In this section, we present our findings for RQ\textsubscript{1} and RQ\textsubscript{2}. 

\subsection{Issue Frequency \& Distribution (RQ\textsubscript{1})} \label{RQ1}

Table~\ref{tab:agent_statistics} summarizes the overall distribution of SonarQube issues observed in merged agent-generated bug-fix PRs. In absolute terms, OpenAI Codex contributes the most issues (456), while Claude Code contributes the fewest (69). However, this pattern largely reflects differences in PR volume rather than intrinsic differences in code quality, as Codex contributes the most fix PRs, while Claude Code contributes the fewest. 
%
Nevertheless, some contrasts emerge when comparing agents with more comparable PR counts. For example, Copilot and Devin contribute a similar number of fix PRs (106 and 100, respectively), yet Copilot is associated with more than twice the total number of issues (235 vs. 99) and a higher mean issue count per PR (2.2 vs. 1.0). Cursor exhibits a more pronounced imbalance: despite contributing only 40 fix PRs, it accounts for the second-highest total number of issues (331), along with the highest average issues per PR (mean = 8.3) and substantial dispersion (std = 38.2, max = 233). These observations suggest that cumulative issue counts alone can be misleading when sample sizes differ markedly across agents.
Estimated technical debt follows a similar pattern, with higher cumulative debt primarily driven by contribution volume and change size rather than per-PR quality alone.
 
To mitigate the effect of PR volume, we next examine issues normalized by code churn. As shown in Fig.~\ref{fig:issue_density}, median issue density measured as issues per KLOC is broadly similar across most agents, and a Kruskal–Wallis test indicates no statistically significant differences. This suggests that the apparent disparity in cumulative issue counts is driven by larger code changes rather than systematically poorer code quality. Cursor remains a partial exception, exhibiting a higher median issue density even after normalization, indicating that its fix PRs tend to introduce more issues per unit of code change. Overall, these results indicate that agent-level differences in observed issue frequency are largely explained by contribution volume and change size, reinforcing the need to interpret per-agent breakdowns with caution under imbalanced sampling.
\vspace{-0.4cm}
\begin{center} 
{\setlength{\fboxsep}{6pt}
\colorbox{teal!5!white}{%
  \parbox{0.98\linewidth}{%
    \textbf{RQ\textsubscript{1} Summary:} Agent-generated bug-fix PRs frequently introduce post-merge code quality issues. Observed per-agent differences are largely driven by variation in code change volume.
  }%
}}
\end{center}
\subsection{Severity Profiles \& Prevalence (RQ\textsubscript{2})} \label{sub-sec: RQ2}
While RQ1 focused on frequency distribution and density of the issues, this research question shifts the attention to the severity profiles and prevalence of issue types introduced in agentic PRs.

\textbf{Severity Profiles:} 
Table~\ref{tab:hierarchical_issue_distribution} summarizes the distribution of issue types and severity levels observed in merged agent-generated bug-fix PRs. Across the dataset, \textit{Code Smell} issues dominate, indicating that agentic PRs more often introduce maintainability concerns than functional defects.
While differences across agents are observable, they largely mirror the underlying imbalance in PR contributions. OpenAI Codex accounts for the largest number of \textit{Code Smell} issues, which aligns with its dominant share of fix PRs in the dataset. Cursor, in contrast, contributes a disproportionately high number of \textit{Code Smell} issues despite contributing only 40 fix PRs, while Copilot exhibits a moderate number of issues relative to a substantially larger PR volume. 
Claude Code and Devin introduce fewer \textit{Code Smell} issues for different reasons: Claude Code contributes fewer fix PRs, while Devin contributes a comparable number of PRs but introduces fewer code smells per PR. Across agents, code smells are primarily classified as \texttt{CRITICAL} or \texttt{MAJOR}, indicating that introduced maintainability issues are often non-trivial.

\begin{table}[!ht]
\centering
\caption{Distribution of Issues by Type, Severity, and Agent}
\label{tab:hierarchical_issue_distribution}
\resizebox{\linewidth}{!}{%
\begin{tabular}{llrrrrr}
    \toprule
    \makecell{\textbf{Issue}\\\textbf{Type}} & \textbf{Severity} & \makecell{\textbf{Claude}\\\textbf{Code}} & \textbf{Copilot} & \textbf{Cursor} & \textbf{Devin} & \makecell{\textbf{OpenAI}\\\textbf{Codex}} \\
    \midrule
            Bug         & BLOCKER & 16 & 4 & 0 & 1 & 20 \\
                        & MAJOR & 1 & 0 & 0 & 2 & 4 \\
            \midrule
            \textbf{Total} &  & \textbf{17} & \textbf{4} & \textbf{0} & \textbf{3} & \textbf{24} \\
            \midrule
            Code Smell  & BLOCKER & 0 & 0 & 2 & 0 & 1 \\
                        & CRITICAL & 28 & 166 & 129 & 45 & 162 \\
                        & INFO & 2 & 16 & 8 & 3 & 66 \\
                        & MAJOR & 12 & 31 & 124 & 10 & 98 \\
                        & MINOR & 3 & 7 & 66 & 21 & 59 \\
            \midrule
            \textbf{Total} &  & \textbf{45} & \textbf{220} & \textbf{329} & \textbf{79} & \textbf{386} \\
    \midrule
    Security Hotspot    & HIGH & 1 & 0 & 1 & 0 & 5 \\
                        & LOW & 6 & 10 & 0 & 17 & 26 \\
                        & MEDIUM & 0 & 1 & 1 & 0 & 15 \\
                        \midrule
    \textbf{Total} &  & \textbf{7} & \textbf{11} & \textbf{2} & \textbf{17} & \textbf{46} \\
    \bottomrule
\end{tabular}%
}
\vspace{-0.3cm}
\end{table}

\textit{Bug} and \textit{Security Hotspot} issues occur less frequently than \textit{Code Smells} but show uneven distributions across agents.OpenAI Codex contributes the most detected bugs, largely classified as \texttt{BLOCKER}, consistent with its dominant share of bug-fix PRs. Conversely, Claude Code contributes the second-highest number of bug issues despite a small PR volume, indicating a higher concentration of severe defects. A similar pattern is observed for \textit{Security Hotspots}. OpenAI Codex contributes the largest number of hotspots, including the highest number of \texttt{HIGH}-severity cases, again aligning with its PR volume, while other agents introduce fewer hotspots that are predominantly \texttt{LOW} severity. No \textit{Vulnerability} issues are detected, and \textit{Duplicate LOC} remains below 3\% across all agents.

\textbf{Prevalence: }To further understand the nature of the detected quality issues, we analyzed the specific SonarQube rules violated by agentic PRs.
For \textit{Bugs}, violations are relatively limited in number but correspond to concrete correctness problems. The most frequently violated rule, \textcolor{Mycolor}{\textit{python:S930}} (incorrect number of function arguments), accounts for nearly half of all detected bug issues, with the majority of these violations originating from Claude-generated code. Other bug-related violations, such as misuse of non-iterables in loops \textcolor{Mycolor}{(\textit{python:S3862})} and identical implementations in conditional branches \textcolor{Mycolor}{(\textit{python:S3923})}, occur less frequently but still point to lapses in logical correctness that are not caught prior to merge.
\begin{table*}[t]
\centering
\caption{Top 3 Most Violated SonarQube Rules per Issue Type with Agent Distribution}
\label{tab:top5_rules_per_type}
\small
\begin{tabularx}{\textwidth}{@{}l p{1.6cm} X *{6}{>{\centering\arraybackslash}c}@{}}
    \toprule
    \textbf{Issue Type} &
    \textbf{Rule ID} &
    \textbf{Description} &
    \textbf{Total} &
    \textbf{Claude} &
    \textbf{Copilot} &
    \textbf{Cursor} &
    \textbf{Devin} &
    \textbf{Codex} \\
    \midrule
    \multirow{3}{*}{Bug}
      & S930  & Function calls should use the correct number of arguments
      & 23 & 16 & 0 & 0 & 0 & 7 \\
      & S3862 & ``for of'' should not be used with non-iterables
      & 15 & 0 & 4 & 0 & 0 & 11 \\
      & S3923 & Avoid identical implementations across conditional branches
      & 3 & 0 & 0 & 0 & 0 & 3 \\
    \midrule
    \multirow{3}{*}{Code Smell}
      & S1192 & String literals should not be duplicated
      & 212 & 22 & 69 & 25 & 33 & 63 \\
      & S3776 & Cognitive complexity of functions should not be too high
      & 157 & 6 & 25 & 53 & 11 & 62 \\
      & S1172 & Unused function parameters should be removed
      & 114 & 0 & 6 & 75 & 2 & 31 \\
    \midrule
    \multirow{3}{*}{Security Hotspot}
      & Encrypt-data & Encrypting data is security-sensitive
      & 33 & 0 & 10 & 0 & 7 & 16 \\
      & Others & Using publicly writable directories is security-sensitive
      & 22 & 6 & 0 & 0 & 10 & 6 \\
      & Weak-cryptography & Use of weak cryptographic mechanisms is security-sensitive
      & 17 & 0 & 1 & 1 & 0 & 15 \\
    \bottomrule
\end{tabularx}
\end{table*}
In contrast, \textit{Code Smell} violations are substantially more frequent and span a wider range of maintainability concerns. The most commonly violated rule is \textcolor{Mycolor}{\textit{python:S1192}} (duplicated string laterals), followed by \textcolor{Mycolor}{\textit{python:S3776}} (excessive cognitive complexity) and \textcolor{Mycolor}{\textit{python:S1172}} (unused function parameters). These violations suggest that agentic code often introduces redundancy and complexity that may hinder readability and long-term maintenance.
Additionally, \textit{Security Hotspots} occur less frequently but highlight sensitive security-related violations in agent-generated code. The most commonly flagged rule concerns \textcolor{Mycolor}{\textit{encryption-related}} operations, followed by the use of \textcolor{Mycolor}{\textit{publicly writable directories}} and \textcolor{Mycolor}{\textit{weak cryptographic}} mechanisms. Although these findings do not necessarily indicate exploitable vulnerabilities, they suggest that agent-generated PRs may introduce security concerns that warrant closer inspection during code review.
\vspace{-0.4cm}
\begin{center} 
{\setlength{\fboxsep}{6pt}
\colorbox{teal!5!white}{%
  \parbox{0.98\linewidth}{%
    \textbf{RQ\textsubscript{2} Summary:} \textit{Code Smells} are prevalent across merged agent-generated bug-fix PRs at all severity levels, whereas \textit{Bugs} appear less frequently but are often classified as blockers.
  }%
}}
\end{center}

\section{Key Findings and Suggestions} \label{Sec:Discussion}
\textbf{Efficiency–Quality Trade-offs in Agentic PRs.} Across agent-generated bug-fix PRs, faster merge outcomes and higher acceptance rates do not consistently align with post-merge code quality. Although agents contributing more PRs account for a larger share of introduced \textit{Code Smells} and \textit{Security Hotspots}, these patterns are largely driven by PR volume and code churn rather than intrinsic agent-specific quality differences. This suggests a broader efficiency–quality trade-off in agentic PR integration. Consistent with prior evidence that agentic bug-fix PRs pose trust and review challenges \cite{watanabe2025use, rondon2025evaluating}, our findings indicate that merge success alone is insufficient as a quality signal. Thus, projects should adopt systematic, size-aware quality checks beyond unit tests \cite{chen2025evaluating}, including static analysis gates and targeted security review, for agent-generated PRs merged at scale.

\textbf{The `Maintainability Debt' of AI Agents:} The frequent violations of duplicated literals (\textit{S1192}) and high congnitive complexity (\textit{S3776}) indicate that agentic fixes often introduce maintainability smells. Prior studies show that code smells are associated with higher change-proneness and maintenance effort and often persist once introduced \cite{tufano2017and, palomba2018diffuseness}. Cognitive complexity was also introduced to capture code understandability costs beyond traditional complexity metrics \cite{Barn2020AnEVA}. Teams adopting agentic workflows should treat duplication and high complexity as review triggers and enforce maintainability gates alongside testing. Otherwise, technical debt may accumulate through incremental merges \cite{tufano2017and}.

\textbf{Rare-but-Severe Bug ``Reliability Paradox''.} Although \textit{Bugs} are less frequent than other issues in the bug-fix PRs, they are disproportionately severe (often \texttt{BLOCKER}), which are likely to impact runtime behaviour in production \cite{sonarqube-docs}. For example, \textit{python:S930} (incorrect number of function arguments), which can directly cause runtime \textit{TypeError}s due to mismatches with existing API contracts \cite{sonarqube-rule}. 
This aligns with prior findings that LLM-generated code often misuses APIs or parameters, producing syntactically valid but semantically misaligned changes~\cite{zhuo2025identifying}. Accordingly, reviews of agentic bug-fix PRs should prioritize correctness checks and require full CI execution before merging.
\vspace{-0.3cm}
\section{Related Works} \label{Sec:Related-works}
Recent studies have explored the quality of agentic code in both controlled and realistic settings. Nikolaidis et al. evaluated ChatGPT (v3.5) and GitHub Copilot on 60 LeetCode problems in Python, analyzing correctness as well as code-quality proxies (e.g., cyclomatic complexity, token count, lines of code). They observed that iterative “improvement” is not consistently monotonic, reinforcing the need for human validation \cite{nikolaidis2024chatgptcopilot}. Yetistiren et al. \cite{yetistiren2022assessing} compared multiple code generation tools using various code quality metrics and found that Copilot was able to generate syntactically valid code with a 91.5\% success rate. In contrast, Fu et al. \cite{fu2025security} analyzed GitHub Copilot–generated code and reported a high likelihood of quality issues related to security weaknesses. Chen and Jiang \cite{chen2025evaluating} studied software development agents on SWE-bench Verified issues and assessed patch characteristics and post-patch quality signals using static analysis; they report that agents vary in effectiveness and that quality outcomes differ across agents

While prior work performed benchmark evaluations and patch studies, we fill a critical gap by analyzing merged bug-fix PRs from real GitHub repositories in the AIDev dataset and characterizing newly introduced issues using differential SonarQube analysis.
\vspace{-0.3cm}
\section{Threats to Validity} \label{Sec:Threats}
Our findings are limited to merged, agent-generated bug-fix PRs in Python repositories from the AIDev dataset \cite{li2025aidev} and may not generalize to other languages or PR intents.
Our study lacks a matched human baseline and focuses on static post-merge quality signals, excluding downstream runtime and maintenance effects.
Moreover, agent representation in the dataset is highly imbalanced, with some agents appearing infrequently, thereby reducing statistical power. Accordingly, agent-level results are exploratory and interpreted with caution for agents with small sample sizes. 
We therefore report per-agent sample sizes to avoid over-interpreting small samples. Finally, SonarQube’s tool-defined severity labels and the use of the Community Edition may not fully reflect production risk. To mitigate this, we kept the analysis toolchain fixed and fully documented, and we framed findings under the chosen setup.


\section{Conclusion} \label{Sec:Conclusion}
This study presents a large-scale empirical analysis of merged agentic bug-fix pull requests, revealing that apparent differences in code quality across agents are largely explained by PR size. Across all agents, maintainability-related issues dominate, while functional bugs are less frequent but severe, and security-relevant patterns appear unevenly across agents. These findings show that merge success and rapid integration are unreliable indicators of code quality, motivating systematic, size-aware quality checks and review support for agent-generated contributions.

\section{Acknowledgment}
This research is supported in part by the Natural Sciences and Engineering Research Council of Canada (NSERC) Discovery Grants program, the Canada Foundation for Innovation's John R. Evans Leaders Fund (CFI-JELF), and by the industry-stream NSERC CREATE in Software Analytics Research (SOAR).

\bibliographystyle{ACM-Reference-Format}
\bibliography{MANUSCRIPT}
\end{document}